# Status And Highlights Of VERITAS

Nicola Galante[a] for the VERITAS Collaboration[b]

[a]*Harvard-Smithsonian Center for Astrophysics*
*ngalante@cfa.harvard.edu*
[b]*http://veritas.sao.arizona.edu*

**Abstract.** VERITAS (Very Energetic Radiation Imaging Telescope Array System) is an array of atmospheric Cherenkov telescopes sensitive to very high energy (VHE) γ-rays above 100 GeV. Located at the Fred Lawrence Whipple Observatory in southern Arizona, USA, the VERITAS array of four 12m-diameter telescopes began full operation in September 2007. Two major upgrades, the relocation of telescope 1 in Summer 2009 and the upgrade of the level-2 trigger in Fall 2011, made VERITAS the most sensitive VHE instrument in the northern hemisphere. The VERITAS Collaboration consists of scientists from institutions in the USA, Canada, Germany and Ireland. VERITAS is performing observations that cover a broad range of science topics, including the study of galactic and extragalactic astrophysical sources of VHE radiation and the study of particle astrophysics, such as the indirect search for dark matter in astrophysical environments. The VERITAS observational campaigns resulted in the detection of 40 VHE sources, including the discovery of 20 new VHE γ-ray emitting sources. Here we summarize the current status of the observatory, describe the recent scientific highlights and outline plans for the future.

## THE VERITAS OBSERVATORY

The VERITAS (Very Energetic Radiation Imaging Telescope Array System) detector is an array of four 12-m diameter imaging atmospheric-Cherenkov telescopes located in southern Arizona [1]. The optical design of each telescope is the traditional Davies-Cotton with an f/D=1 and a segmented reflective surface of 350 hexagonal aluminum-coated glass mirrors. Each telescope is equipped with a camera consisting of 499 photo-multiplier tubes (PMT), covering a ~3.5° field of view (FoV). Main parts of the electronic chain are a 3-level topological trigger and a flash analog digital converter (FADC) data acquisition system (DAQ) capable of recording events at 500 MSample/s with a typical dead time <10%. Designed to detect emission from astrophysical objects in the energy range from 100 GeV to greater than 30 TeV, VERITAS has an energy resolution of ~15% and an angular resolution (68% containment) of ~0.1° per event at 1 TeV. VERITAS records more than 1000 hours of data every year and is capable of operating under moderate Moon-light conditions. Approximatively 25% of the data recorded by VERITAS are taken under Moon-light conditions.

## Upgrades 2009-2011

In Summer 2009 one of the four telescopes was relocated to a different position, increasing the overall sensitivity of the array by about 30% [2]. A source with a flux of 1% of the Crab Nebula flux is detected in less than 30 hours of observations, while a 5% Crab Nebula flux source is detected in less than 2 hours.

A second upgrade of VERITAS occurred in November 2011, when the level-2 trigger was replaced by a new topological trigger capable of identifying air shower events on a shorter coincidence time scale. This reduces the contamination of the triggered event population from spurious night sky background (NSB) illumination.

## Upgrade 2012

In Summer 2012 VERITAS replaced the old Photonis XP2970 PMTs of the camera in each telescope with new super-bialkali PMTs (Hamamatsu R10560-100-20) characterized by higher quantum efficiency (QE) and shorter time profile. The new PMTs specifications, together with the previously installed topological trigger, are expected to increase VERITAS efficiency in detecting Cherenkov light. VERITAS has already commissioned the new cameras and is currently operating full duty.

## VERITAS SCIENCE

The long-term VERITAS Science Plan is dedicated to the study of the physical processes responsible for γ-ray emission in a large variety of sources and astrophysical environments. The plan is based on four major themes of scientific exploration:

1. Particle Physics and Fundamental Laws (search for dark matter particle candidates, Lorentz invariance violation, evaporating primordial black holes, search for axion-like particles).
2. Cosmology (γ-ray opacity measurements, cosmic rays in starburst galaxies and ultra-luminous infrared galaxies, large-scale structures and intergalactic magnetic fields).
3. Black holes and acceleration processes in their surroundings.
4. Physical processes in galactic tevatrons and pevatrons.

## The VERITAS Observation Strategy

VERITAS observations are almost entirely based on proposal of observations submitted by VERITAS members and by external scientists through a VERITAS member. Only ~10% of VERITAS observation time is dedicated to discretionary director time or automatic target of opportunity (ToO) observations, namely gamma-ray bursts (GRB). In average, more than 90% of VERITAS data are taken with the array fully operative and under good weather conditions. The current VERITAS catalog, resulting from five years of VERITAS observations, can be seen in figure 1.

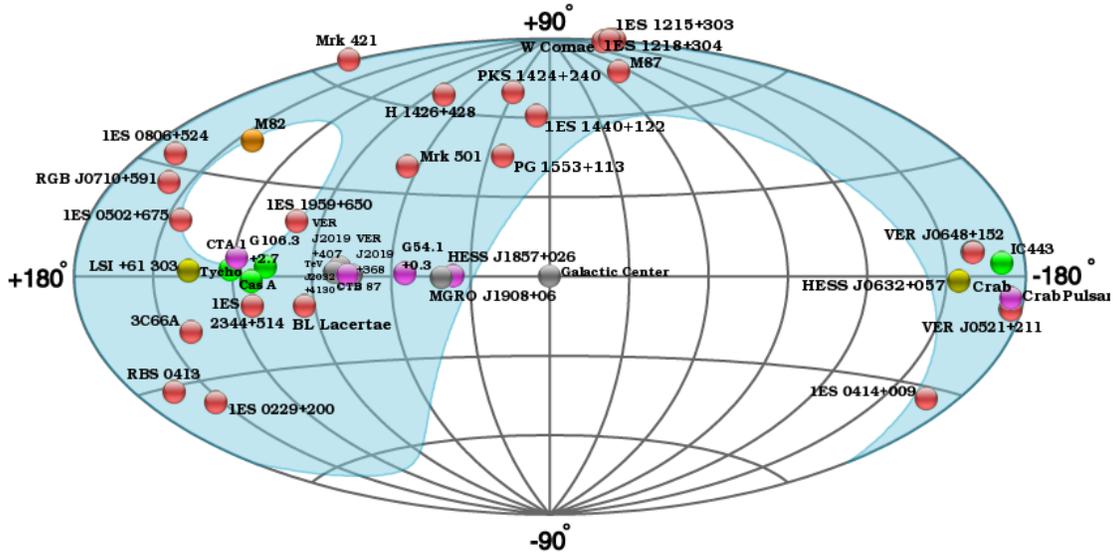

**FIGURE 1.** The VERITAS catalog at the time of these proceedings. It includes 40 γ-ray sources: 21 blazars, 12 galactic sources, 1 radio galaxy, 1 starburst galaxy and 5 unidentified γ-ray emitters. Of these 40 sources, 21 have been discovered as TeV emitters by VERITAS. *Red*: active galactic nuclei. *Copper*: starburst galaxy. *Magenta*: pulsar wind nebulae. *Green*: shell remnants, supernova remnants, molecular clouds. *Gold*: X-ray binaries, pulsars. *Grey*: unidentified, other.

## The VERITAS Galactic Program

### *Binary Systems*

Gamma-ray binaries constitute a small class of galactic objects characterized by variable non-thermal emission across all wavelengths, in most cases modulated by the orbital motion of an O or B star and a compact object, as both the massive star and the compact object orbit the center of mass.

The γ-ray binary LS I +61 303 consists of a Be star and a compact object of unknown nature. The underlying acceleration and gamma-ray emission scenario is actively discussed in literature, with several inconclusive observations. VERITAS is monitoring it since 2006, covering its entire 26.5 days orbit and revealing cumulative emission on both the apastron and periastron phases. This source is not detected by VERITAS every year and daily variability is seen. It is still unclear if the observed quasi-periodic emission is due to a multi-year modulation or to some stochastic process overlapped to the orbital modulation.

γ-ray emission from HESS J0632+057 above 400 GeV was serendipitously discovered during observations of the Monoceros Loop Supernova Remnant in 2004/2005 [3]. Follow-up observations by VERITAS revealed a variable gamma-ray source [4], coincident with the hard X-ray source XMMU J063259.3+054801 and the Be star MWC 148. Deep observations of HESS J0632+057 with VERITAS over several years revealed an apparently predictable light curve with maximum VHE emission simultaneous with the maximum in X-ray emission.

### *Crab Pulsar*

The spectrum of pulsed gamma rays from pulsars γ-ray data shows a spectral cutoff above a few GeV [5]. Traditional pulsar models attribute this cutoff to curvature radiation within the magnetosphere. VERITAS collected 107 hours of low-zenith observations from 2007 to 2011. Significant signal is discovered by VERITAS above 100 GeV up to 400 GeV, incompatible with the expectation of an exponential cutoff from models based on curvature radiation processes.

### *SNR*

CTA1 is a composite type SNR, observed with shell type morphology in radio wavelengths and center filled morphology in X-rays. In addition to the diffuse emission extending from the center region to the radio shell, X-ray observation revealed a point source (RX J0007.0+7302) and a compact pulsar wind nebular (PWN) with a jet-like structure. VERITAS observed the CTA1 region for two seasons from 2010 to 2012, collecting a total of ~41 hours of exposure time. A TeV source was found with a maximum significance of 6.5σ. The centroid of the VERITAS source is located ~5 arcmin away from PSR J0007+7303. The properties of CTA1 agree well with the TeV/X-ray PWN population study by Kargaltsev and Pavlov [6], supporting the PWN origin of the TeV emission.

## The VERITAS AGN PROGRAM

Since September 2007, VERITAS has observed 128 blazars, for a total of ~2000 h (annual average of ~400 h) during good weather conditions. Until 2010, ~80% of the VERITAS blazar data were used for VHE discovery observations and follow-up studies of any new sources. In 2010, the focus became deep studies of known VHE sources, and ~60% of the data is now devoted to known sources. Target-of-Opportunity (ToO) observations are a key part of the blazar program and typically average 50 h per year.

The VERITAS AGN science program develops along four major directions:

• **Discovery program:** create a VHE AGN catalog, discover new AGN types and identify individual and class features though population studies.

• **Multi-wavelength (MWL) campaigns:** deep understanding of individual sources through correlation studies, time-resolved spectral studies and broadband characterization of the spectral energy distribution (SED).

• **Target of Opportunities:** take advantage of the high-flux state, i.e. higher photon statistics, for accurate measurements of spectral and timing properties.

• **Cosmology:** give the preference to large redshift AGNs in order to investigate physical quantities related to the history of star and galaxy formation in the Universe. The intergalactic magnetic fields (IGMF) and the extragalactic background light (EBL) are connected to the history of the Universe and can be indirectly investigated through γ-ray observation of distant γ-ray sources.

VERITAS observations of AGNs resulted in 22 detections (21 BL-Lac objects and 1 radio galaxy), including 10 TeV discoveries. VERITAS observations of AGNs resulted in the first classification of objects through TeV observations (VER J0521+211/RGB J0521.8+2112 and VER J0648+152/RX J0648.7+1516). VERITAS started the intermediate BL-Lac catalog (IBL) at VHE, detecting 5 IBLs (3C 66A, W Com, PKS 1424+240, B2 1215+30, 1ES 1440+122) and recently detected fast γ-ray variability in BL Lacertae. VERITAS conducts observation within coordinated MWL broadband observational campaigns on the two historical blazars Mrk 421 and Mrk 501. These monitoring campaigns made possible to detect a big flare on Mrk 421 on February 17, 2010, revealing variability on time scales of few minutes, enabling high-precision time-resolved spectral analysis and Lorentz invariance violation studies.

M 87 is the only radio galaxy detected by VERITAS so far. VERITAS conducts extensive MWL campaigns since several years, covering the entire electromagnetic spectrum. The VERITAS-led MWL observational campaign resulted in the first identification of the region responsible for the γ-ray emission in a flaring episode, as the region close to the core.

## The VERITAS DARK MATTER PROGRAM

The indirect search for very high energy (VHE) γ-rays resulting from the annihilation of Dark Matter (DM) particles into standard model particles is an importat contribution of γ-ray astronomy to the particle Physics field. Among theoretical candidates for the DM particle, a weakly interacting massive particle (WIMP) is well-motivated in naturally providing the measured present day cold DM density [7]. Candidates for WIMP dark matter are present in many extensions of the standard model of particle physics, such as supersymmetry (SUSY) [8] or theories with extra dimensions [9]. In such models, the WIMPs either decay or self-annihilate into standard model particles, producing either a continuum of γ-rays with energies up to the DM particle mass, or mono-energetic γ-ray lines.

Promising targets for indirect DM searches are nearby regions with high DM density. The Galactic Center is likely the brightest source of γ-rays resulting from DM annihilations, however the detected VHE γ-ray emission is coincident with the supermassive black hole Sgr A* and a nearby pulsar wind nebula. The complex morphology of the galactic center region requires a dedicated strategy to identify the DM-dominated regions with minimal γ-ray contamination from the pulsar wind nebula emission. No γ-ray emission, established to be related to DM particles annihilation, is detected.

The complexity of the galactic center region also motivates searches for DM annihilation where the VHE γ-ray background is expected to be significantly lower. Satellite dwarf spheroidal galaxies (dSphs) of the Milky Way are attractive targets for indirect DM searches due to their relatively well-constrained DM profiles derived from stellar kinematics, proximity of 20–100 kpc, and general lack of active or recent star formation suggesting a relatively low background from conventional astrophysical VHE processes [10]. VERITAS observed the dSph galaxy Segue 1 between January 2010 and May 2011 for a total of 47.8 hours of quality-selected data [11]. VERITAS observation can constrain expectation from Sommerfeld enhancement effects (boost of the annihilation cross section) and places the strongest limit on the annihilation cross section of the neutralino in dSph galaxies (figure 2).

## ACKNOWLEDGMENTS

This research is supported by grants from the U.S. Department of Energy Office of Science, the U.S. National Science Foundation and the Smithsonian Institution, by NSERC in Canada, by Science Foundation Ireland (SFI 10/RFP/AST2748) and by STFC in the U.K. We acknowledge the excellent work of the technical support staff at the Fred Lawrence Whipple Observatory and at the collaborating institutions in the construction and operation of the instrument.

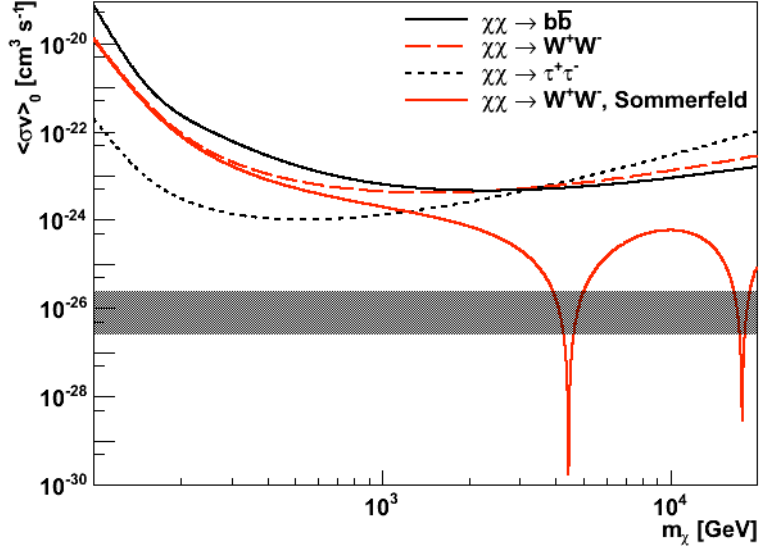

**FIGURE 2.** 95% CL exclusion curves from the VERITAS observations of Segue 1 on ⟨σv⟩/S⁻ as a function of the dark matter particle mass, in the framework of two models with a Sommerfeld enhancement. The expected Sommerfeld enhancement S⁻ applied to the particular case of Segue 1 has been computed assuming a Maxwell dark matter relative velocity distribution. The grey band area represents a range of generic values for the annihilation cross-section in the case of thermally produced dark matter.